\documentclass[aip,reprint]{revtex4-1}
\usepackage{epsfig,amssymb,amsmath}

\draft 

\begin{document}


\title{Magnetization reversal by superconducting current in $\varphi_0$ Josephson junctions} 




\author{Yu. M. Shukrinov}
\affiliation{BLTP, JINR, Dubna, Moscow Region, 141980, Russia}
\affiliation{Dubna State University, Dubna,  141980, Russia}

\author{I. R. Rahmonov}
\affiliation{BLTP, JINR, Dubna, Moscow Region, 141980, Russia}
\affiliation{Umarov Physical Technical Institute, TAS, Dushanbe, 734063, Tajikistan}

\author{K. Sengupta}
\affiliation{Theoretical Physics Department, Indian Association for the Cultivation of Science, Jadavpur, Kolkata 700032, India}

\author{A. Buzdin}
\affiliation{University Bordeaux, LOMA UMR-CNRS 5798, F-33405 Talence Cedex, France}


\date{\today}

\begin{abstract}
We study magnetization reversal in a $\varphi_0$ Josephson junction with direct coupling between magnetic moment
and Josephson current. Our simulations  of magnetic moment dynamics show that by applying an electric current pulse, we
can realize the full magnetization reversal. We propose different protocols of full magnetization reversal based on the variation of the Josephson junction and pulse parameters, particularly, electric current pulse amplitude, damping of magnetization and spin-orbit interaction. We discuss experiments which can probe the magnetization reversal in $\varphi_0$-junctions.
\end{abstract}

\maketitle
Spintronics, which deals with an active control of spin dynamics
in solid state systems, is one of the most rapidly developing field
of condensed matter physics \cite{sdsrev}. An important place in
this field is occupied by superconducting spintronics dealing with
the Josephson junctions (JJ) coupled to  magnetic systems
\cite{linder15}. The possibility of achieving electric control over
the magnetic properties of the magnet via Josephson current and its
counterpart, {\it i.e.}, achieving magnetic control over Josephson
current, recently attracted a lot of attention
\cite{buzdin05,bergeret05,golubov04}. Spin-orbit coupling plays a
major role in achieving such control. For example, in
superconductor/ferromagnet/superconductor (S/F/S) JJs, its presence
in a ferromagnet without inversion symmetry provides a mechanism for
a direct (linear) coupling between the magnetic moment and the
superconducting current. In such junctions, called hereafter
$\varphi_0$-junction, time reversal symmetry is broken, and the
current--phase relation is given by $I = I_c \sin
(\varphi-\varphi_0)$, where the phase shift $\varphi_0$ is
proportional to the magnetic moment perpendicular to the gradient of
the asymmetric spin-orbit potential, and also to the applied current.\cite{buzdin08,krive05,reynoso08} . Thus such JJs
allow one to manipulate the internal magnetic moment by Josephson
current \cite{buzdin08,konschelle09}. The static properties of
$S/F/S$ structures are well studied both theoretically and
experimentally; however, the magnetic dynamics of these systems has
not been studied in detail beyond a few theoretical works
\cite{buzdin08,konschelle09,waintal02,braude08,linder83,cai10,chud2016}.

The spin dynamics associated with such $\varphi_0$-junctions was
studied theoretically in Ref.\ \onlinecite{konschelle09}. The
authors considered a $S/F/S$ $\varphi_0$-junction in a low frequency
regime which allowed usage of quasi-static approach to study
magnetization dynamics. It was demonstrated that a DC
superconducting current produces a strong orientation effect on the
magnetic moment of the ferromagnetic layer. Thus application of a DC
voltage to the $\varphi_0$-junction is expected to lead to current
oscillations and consequently magnetic precession. This precession
can be monitored by the appearance of higher harmonics in
current-phase relation; in addition, it also leads to the appearance
of a DC component of the current which increases near a
ferromagnetic resonance\cite{konschelle09}. It is then expected that the presence of external radiation in
such a system would lead to several phenomena such as
appearance of half-integer steps in the current-voltage (I-V)
characteristics of the junction and generation of an additional
magnetic precession with frequency of external radiation
\cite{konschelle09}.

In this paper we study the magnetization reversal in
$\varphi_0$-junction with direct coupling between magnetic moment
and Josephson current and explore
the possibility of electrically controllable magnetization reversal
in these junctions. We carry out investigations of the magnetization dynamics for two types of
applied current pulse: rectangular and Gaussian forms. An exact numerical
simulation of the dynamics of magnetic moment of the ferromagnetic
layer in the presence of such pulses allows us to demonstrate
complete magnetization reversal in these systems. Such reversal
occurs for specific parameters of the junction and the pulse. We chart out these parameters and suggest a possible way for determination of spin-orbit coupling parameter in these systems. We discuss the experiment which can test our theory.

In order to study the dynamics of the S/F/S system, we use the
method developed in Ref.\ \onlinecite{konschelle09}. We assume that
the gradient of the spin-orbit potential is along the easy axis of
magnetization taken to be along ${\hat z}$. The total energy of this
system can be written as
\begin{equation}
E_{\text{tot}}=-\frac{\Phi_{0}}{2\pi}\varphi I+E_{s}\left(  \varphi,\varphi_{0}\right)  +E_{M}\left(  \varphi_{0}\right)  , \label{EQ_energy}%
\end{equation}
where $\varphi$ is the phase difference  between the
superconductors across the junction, $I$ is the external current,
$E_{s}\left( \varphi,\varphi_{0}\right)=E_{J}\left[  1-\cos\left(
\varphi-\varphi_{0}\right)  \right]$, and $\displaystyle
E_{J}=\Phi_{0}I_{c}/2\pi$ is the Josephson energy. Here $\Phi_{0}$
is the flux quantum, $I_{c}$ is the critical current, $\varphi_{0}=l
\upsilon_{so} M_y/(\upsilon_{F} M_{0})$, $\upsilon_{F}$ is Fermi velocity, $l=4 h L/\hbar
\upsilon_{F}$, $L$ is the length of $F$ layer, $h$ is the exchange
field of the $F$ layer, $E_{M}=-K\mathcal{V}M^2_z/(2 M_0^2)$, the
parameter $\upsilon_{so}/\upsilon_{F}$ characterizes a relative
strength of spin-orbit interaction, $K$ is the anisotropic constant,
and $\mathcal{V}$ is the volume of the $F$ layer.

The magnetization dynamics is described by the
Landau-Lifshitz-Gilbert equation\cite{buzdin05} (see  also Supplementary Material) which can be written in the dimensionless form as
\begin{equation}
\label{syseq}
\begin{array}{llll}
 \frac{d m_{x}}{d t}= \frac{1}{1+\alpha^{2}}
\big\{-m_{y}m_{z}+G r m_{z}\sin(\varphi-rm_{y})\\
 - \alpha\big[m_{x}m_{z}^{2}+G r m_{x}m_{y}\sin(\varphi-rm_{y})\big]\big\},
\vspace{0.2 cm}\\
 \frac{d m_{y}}{d t}= \frac{1}{1+\alpha^{2}}
\big\{m_{x}m_{z}\\
 - \alpha\big[m_{y}m_{z}^{2}-G r (m_{z}^{2}+m_{x}^{2})\sin(\varphi-rm_{y})\big]\big\},
\vspace{0.2 cm}\\
 \frac{d m_{z}}{d t}= \frac{1}{1+\alpha^{2}}
\big\{-G r m_{x}\sin(\varphi-rm_{y})\\
 - \alpha\big[G r m_{y}m_{z}\sin(\varphi-rm_{y})-m_{z}(m_{x}^{2}+m_{y}^{2})\big]\big\},
\end{array}
\end{equation}
where $\alpha$ is a
phenomenological Gilbert damping constant, $r=l\upsilon_{so}/\upsilon_{F}$, and $\displaystyle G=
E_{J}/(K \mathcal{V})$. The $m_{x,y,z} = M_{x,y,z}/M_0$ satisfy the constraint
$\sum_{\alpha=x,y,z} m_{\alpha}^2(t)=1$. In this system of equations
time is normalized to the inverse ferromagnetic resonance frequency
$\omega_{F}=\gamma K/M_{0}: (t\rightarrow t \omega_F)$, $\gamma$ is the gyromagnetic ratio, and $M_{0}=\|{\bf M}\|$. In what follows, we obtain time dependence of  magnetization $m_{x,y,z}(t)$, phase difference
$\varphi(t)$ and normalized superconducting current $I_s(t)\equiv
I_s(t)/I_c = \sin(\varphi(t)-rm_{y}(t))$ via numerical solution of Eq.(\ref{syseq}).

\begin{figure}[h!]
 \centering
\includegraphics[height=35mm]{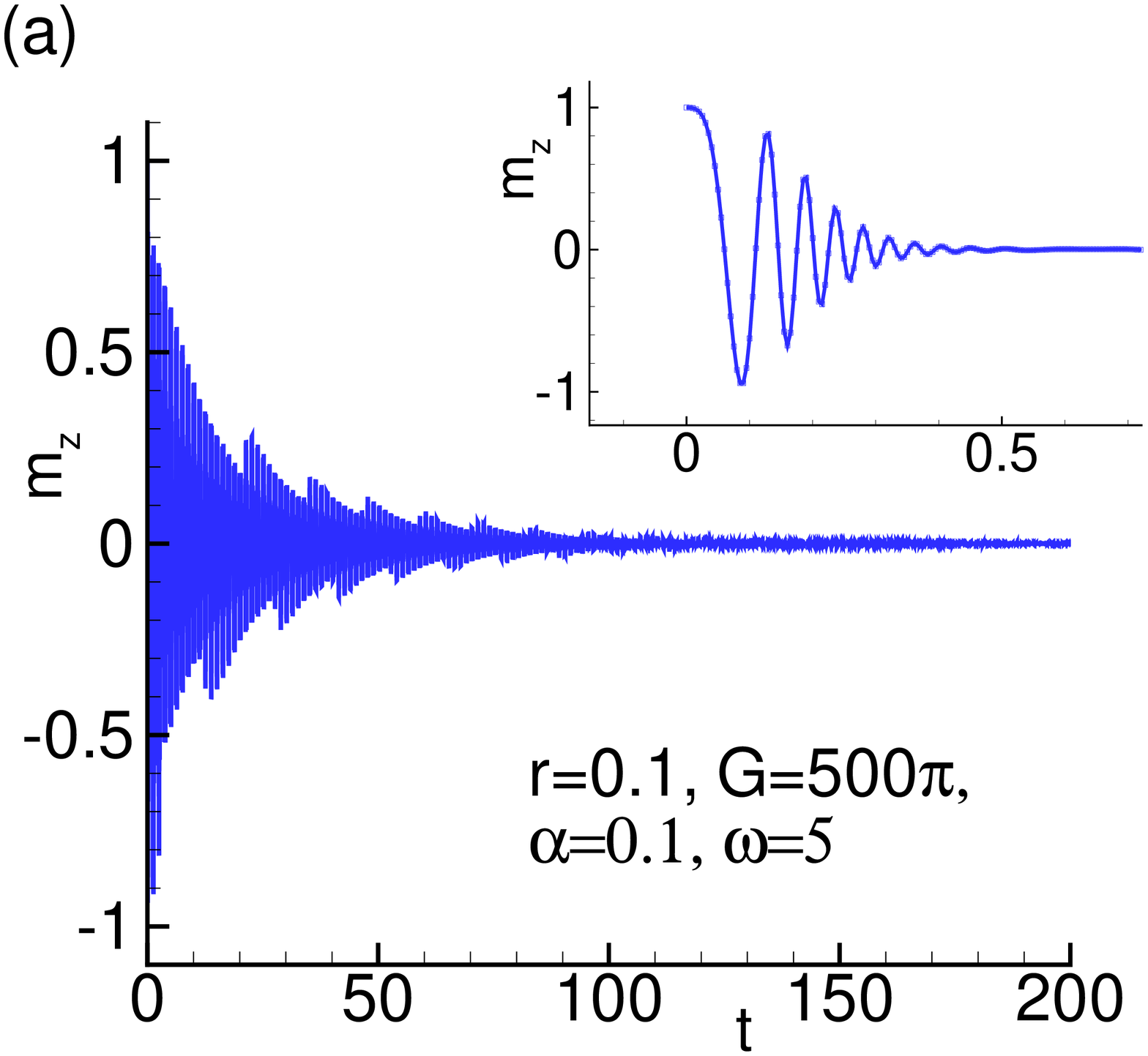}\includegraphics[height=35mm]{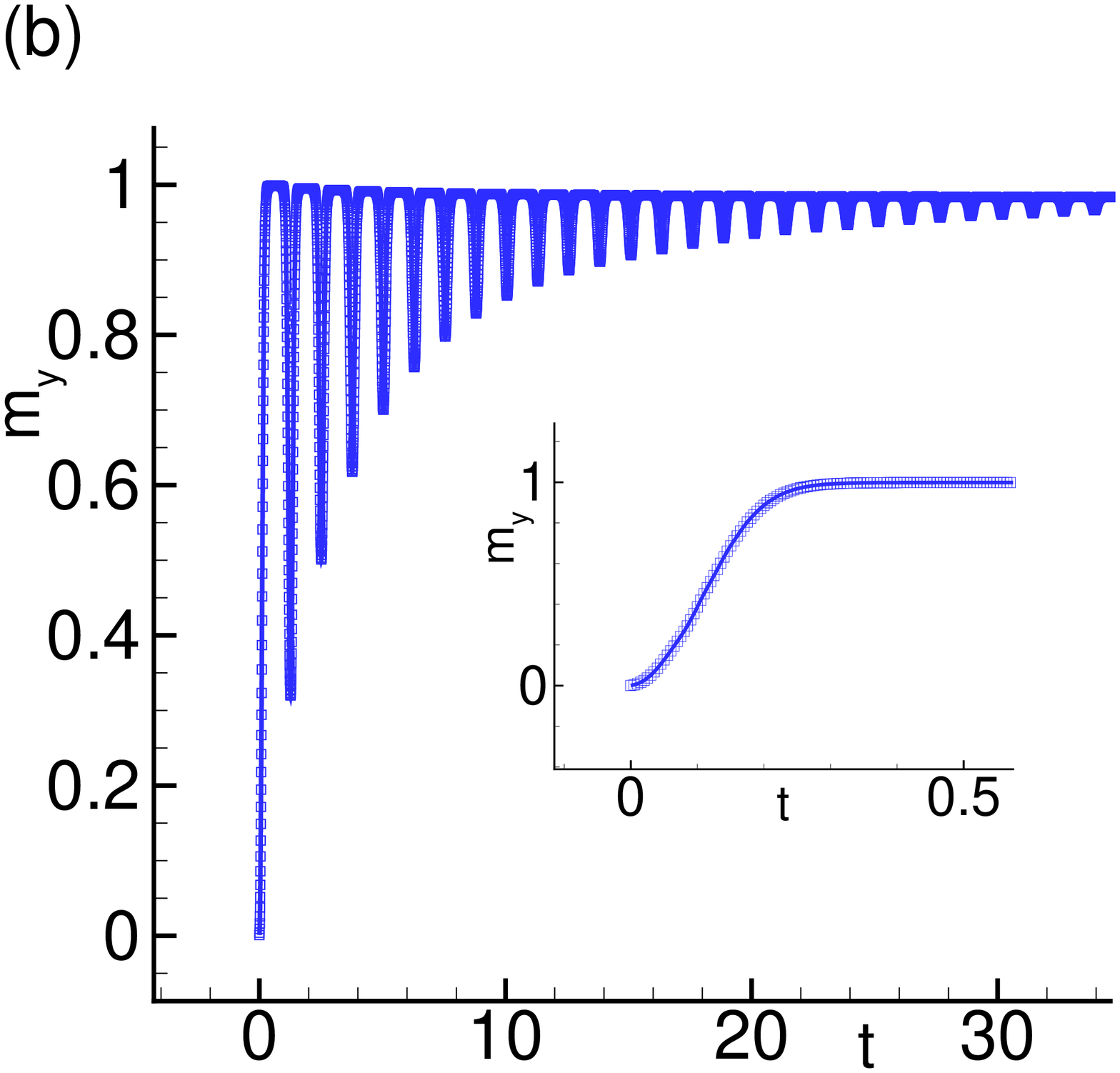}
\caption{(a) Dynamics of $m_{z}$ in case of $\omega_J=5, G=500\pi,r=0.1, \alpha=0.1$. The inset shows the character of time dependence in the beginning  of the time interval; (b) The same as in (a) for $m_y$.}
\label{1}
\end{figure}

Let us first investigate an effect of superconducting current on the dynamics of magnetic momentum. Our main goal is to search for cases related to the  possibility of the full reversal of the magnetic moment by superconducting current.  In Ref.\onlinecite{konschelle09} the authors  have observed a periodic reversal, realized in short time interval. But, as we see in Fig. \ref{1}, during a long time interval the character of $m_z$ dynamics  changes crucially. At long times, $\vec m$ becomes parallel to $y$-axis, as seen from Fig.\ref{1}(b)) demonstrating dynamics of $m_y$. The situation is reminiscent of Kapitza pendulum (a pendulum whose point of suspension vibrates) where the external sinusoidal force can invert the stability position of the pendulum.\cite{kapitza} Detailed features of Kapitza pendulum manifestation will be presented elsewhere.

The question we put here is the following: is it possible to revers the magnetization by the electric current pulse and then preserve this reversed state. The answer may be found by solving the system of equations (\ref{syseq})  together with Josephson relation  $d\varphi/dt=V$, written in the dimensionless form. It was demonstrated in Ref.\onlinecite{cai10} that using a specific time dependence of the bias voltage, applied to the weak link leads to the reversal of the magnetic moment of the nanomagnet. The authors showed the reversal of nanomagnet by linearly decreasing bias voltage $V=1.5-0.00075t$ (see Fig.3 in Ref.\onlinecite{cai10}). The magnetization reversal, in
this case, was accompanied by complex dynamical behavior of the
phase and continued during a sufficiently long time interval.

\begin{figure}[h!]
 \centering
\includegraphics[height=50mm]{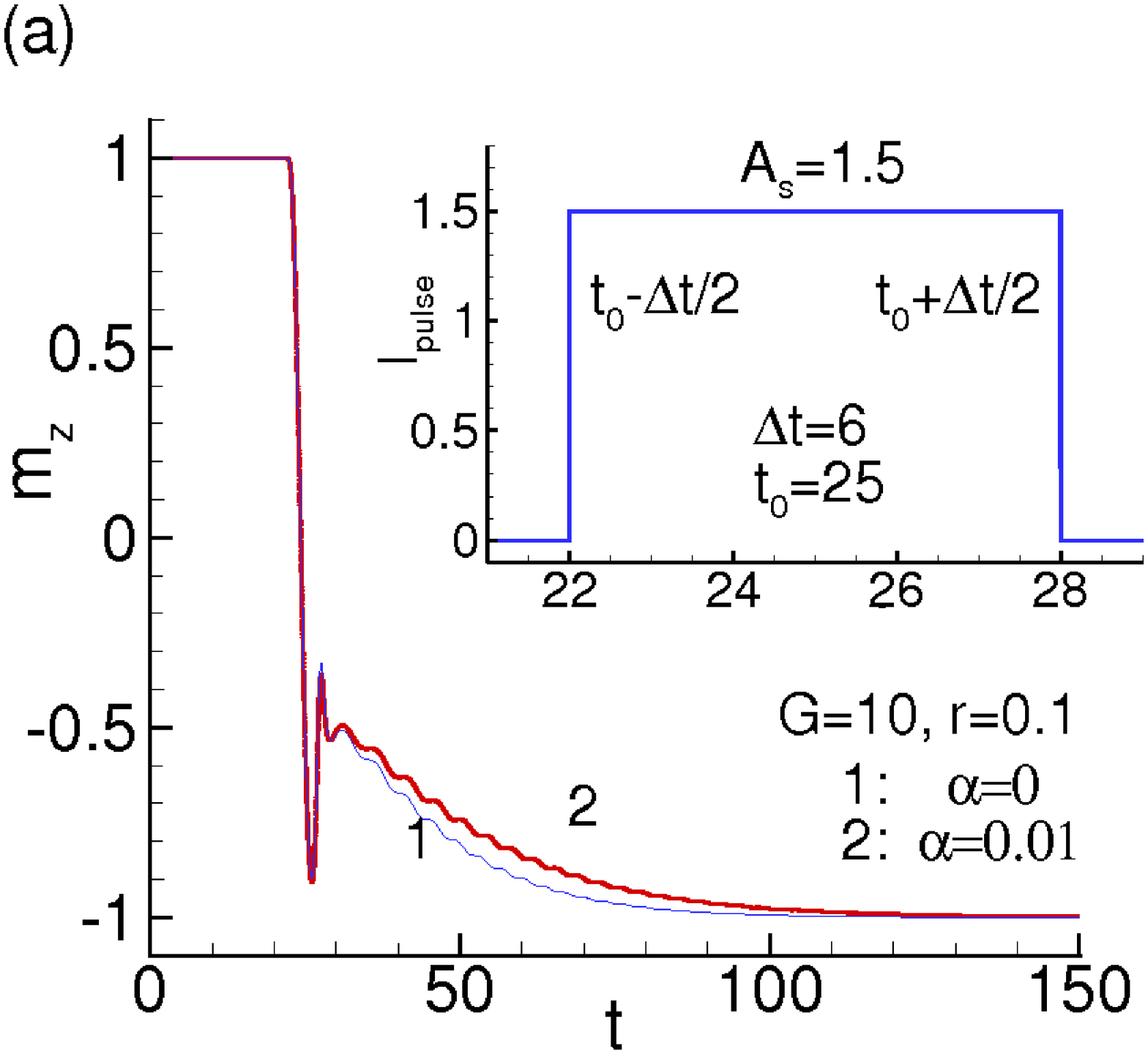}
\includegraphics[height=50mm]{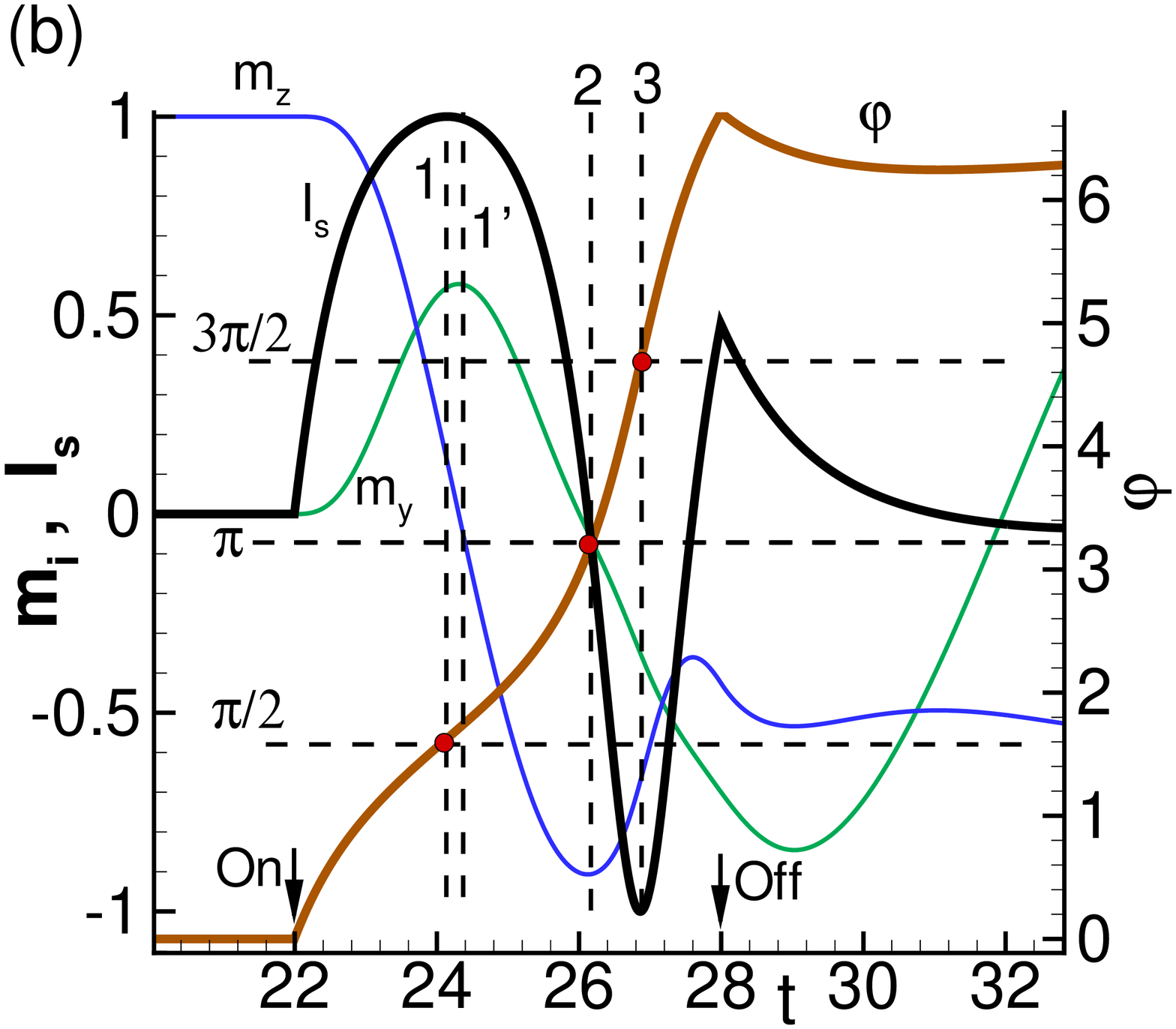}
\caption{
Transition dynamics of the magnetization component $m_z$ for a system with rectangular current pulse shown in the inset; (b) Dynamics of magnetization components together with the phase difference $\varphi$ and superconducting current $I_s$. Arrows indicate the beginning and end of the electric current pulse. Vertical dashed lines indicate the common features while the horizontal ones mark the corresponding values of the phase difference.}
\label{2}
\end{figure}

In contrast, in the present work we investigate the magnetization reversal in the system described by the equations (\ref{syseq}) under the influence of the electric current pulse of rectangular and  Gaussian forms. The effect of rectangular electric current pulse are modeled by  $I_{pulse}=A_{s}$ in the $\Delta t$ time interval ($t_{0}-\frac{\Delta t}{2}, t_{0}+\frac{\Delta t}{2}$) and $I_{pulse}=0$ in other cases. The form of the current pulse is shown in the inset to Fig.\ref{2}(a).

Here we consider the JJ with low capacitance C ( $R^{2}C/L_{J}<<1$, where $L_{J}$ is the inductance of the JJ and $R$ is its resistance), i.e., we do not take into account the displacement current. So, the electric current through JJs is
\begin{equation}
I_{pulse}= w\frac{d\varphi}{dt}+\sin(\varphi-rm_{y})
\label{current}
\end{equation}
where $w=\frac{V_F}{I_cR}=\frac{\omega_F}{\omega_R}$, $V_F=\frac{\hbar\omega_F}{2e}$, $I_c$ - critical current, $R$- resistance of JJ, $\omega_R =\frac{2e I_cR}{\hbar}$ - characteristic frequency.  We solved the system of equations (~\ref{syseq}) together with  equation (~\ref{current}) and describe the dynamics of the system. Time dependence of  the electric current is determined through time dependence of phase difference $\varphi$ and magnetization components $m_x$, $m_y$, $m_z$.

We first study the effect of the rectangular pulse shown in the
inset to Fig.\ref{2}(a). It is found that the reversal of
magnetic moment can indeed be realized  at optimal values of
JJ ($G, r$) and pulse ($A_s, \Delta, t_0$) parameters  .
An example of the transition dynamics for such reversal of $m_{z}$
with residual oscillation is demonstrated in Fig.\ \ref{2}(a);
the corresponding parameter values are shown in the figure.

Dynamics of the magnetic moment components, the phase difference and  superconducting current is illustrated in Fig.\ref{2}(b). We see that in the transition region the phase difference changes from $0$ to $2\pi$ and, correspondingly, the superconducting current changes its direction twice. This is followed by damped oscillation of the superconducting current.
There are some characteristic time points in Fig.\ref{2}(b), indicated by vertical dashed lines. Line 1 corresponds to a phase difference of $\pi/2$ and indicate maximum of superconducting current $I_s$. The line $1^\prime$ which corresponds to the maximum of $m_y$, and  $m_z=0$ has a small shift from line 1. This fact demonstrates that, in general, the characteristic features of $m_{x}$ and $m_{y}$ time dependence do not coincide with the features on the $I_s(t)$, i.e., there is a delay in reaction of magnetic moment to the changes of superconducting current.
Another characteristic point corresponds to the $\varphi=\pi$. At this time line 2  crosses points  $I_s=0$, $m_y=0$, and minimum of $m_z$. At time moment when $\varphi=3\pi/2$ line 3 crosses minimum of $I_s$. When pulse is switched off, the superconducting current starts to flow through the resistance demonstrating damped oscillations and causing residual oscillations of magnetic moment components. Note also, that the time at which the current pulse ends ($t=28$) is actually does not  manifest itself immediately in the  $m_{y}$ (and not shown here $m_{x}$) dynamics. They demonstrate continuous  transition to the damped oscillating behavior.

Fig.\ \ref{2}(b) provides us with a direct way of
determining the spin-orbit coupling strength in the junction via
estimation of $r$. For this, we note that the $\varphi(t) = \varphi_{00} +
\int_0^t V(t') dt'$ can be determined, up to a initial
time-independent constant $\varphi_{00}$, in terms of the voltage
$V(t)$ across the junction. Moreover, the maxima and minima of $I_s$
occurs at times $t_{\rm max}$ and  $t_{\rm min}$ (see Fig.\
\ref{2}(b)) for which $\sin \left[\varphi_{00} + \int_0^{t_{\rm
max}[t_{\rm min}]} V(t') dt' - r m_y(t_{\rm max}[t_{\rm
min}])\right] = +[-]1.$ Eliminating $\varphi_0$ from these
equations, one gets
\begin{eqnarray}
\sin \frac{1}{2} \left[ \int_{t_{\rm max}}^{t_{\rm min}}V(t') dt' +
r [m_y(t_{\rm max})- m_y(t_{\rm min})] \right] = 1 \nonumber\\
\label{mageq1alt}
\end{eqnarray}
which allows us, in principal,  to determine $r$ in terms of the magnetization $m_y$
at the position of maxima and minima of the supercurrent and the
voltage $V$ across the junction. We stress that for the experimental realization of proposed
method one would need to resolve the value of the magnetization at the
time difference of the order of $10^{-10}$ - $10^{-9}$ c. At the present
stage the study of the magnetization dynamics with such a resolution is extremely challenging. To
determine the spin-orbit coupling constant $r$ experimentally it may be more
convenient  to vary the parameters of the current pulse $I(t)$ and study the
threshold of the magnetic moment switching.

\begin{figure}[h!]
 \centering
 \includegraphics[height=35mm]{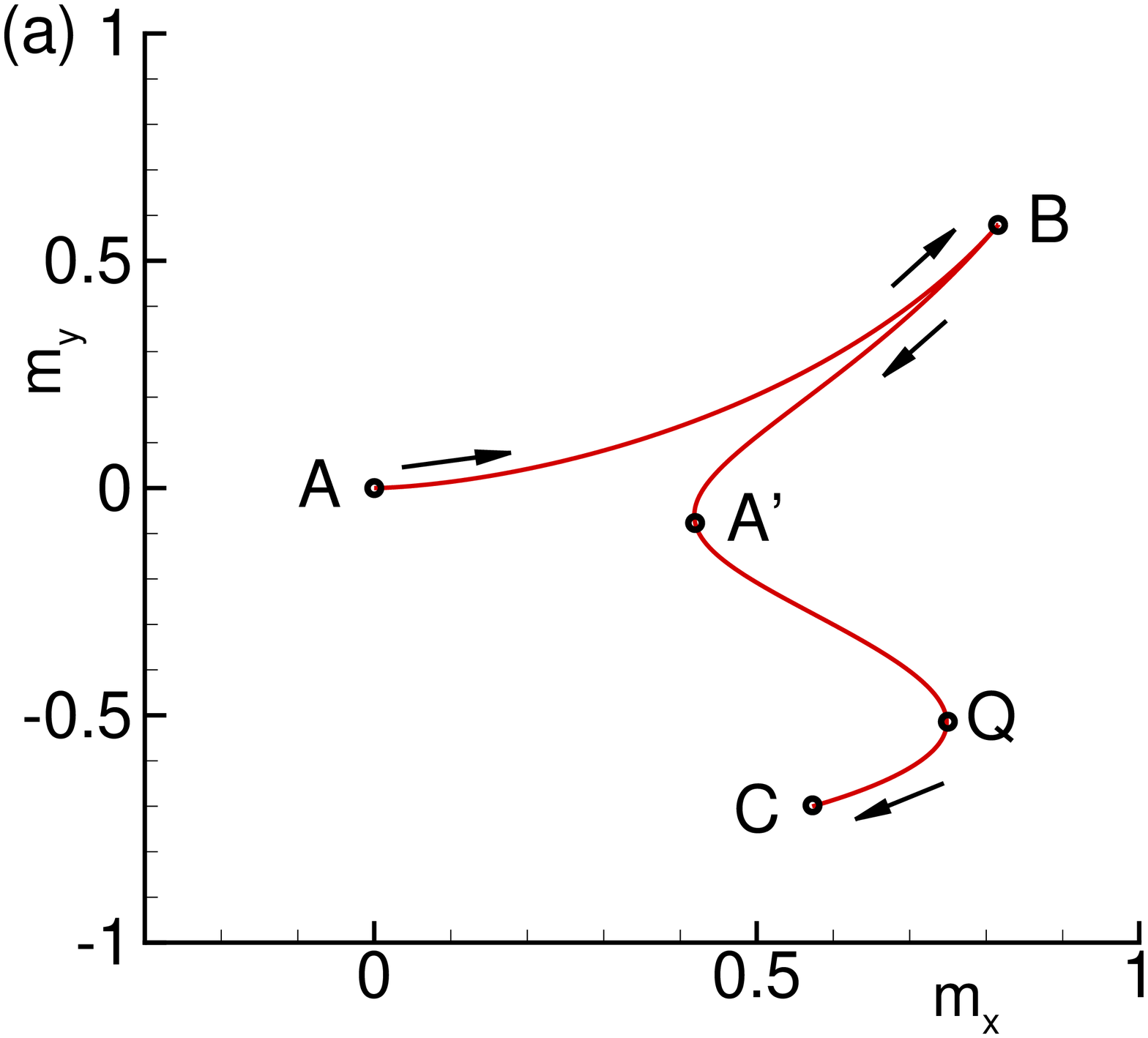}\includegraphics[height=35mm]{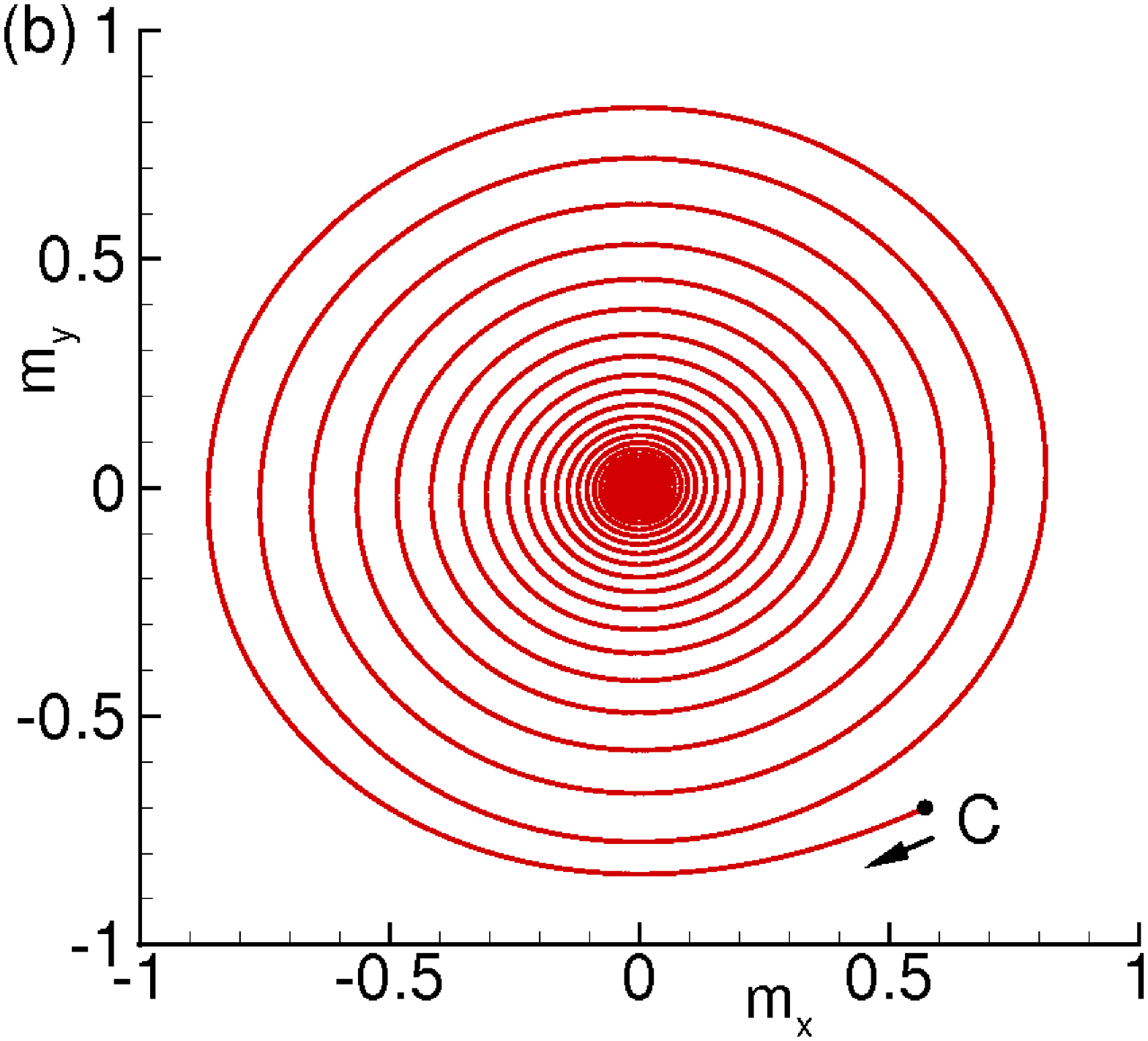}
 \includegraphics[height=35mm]{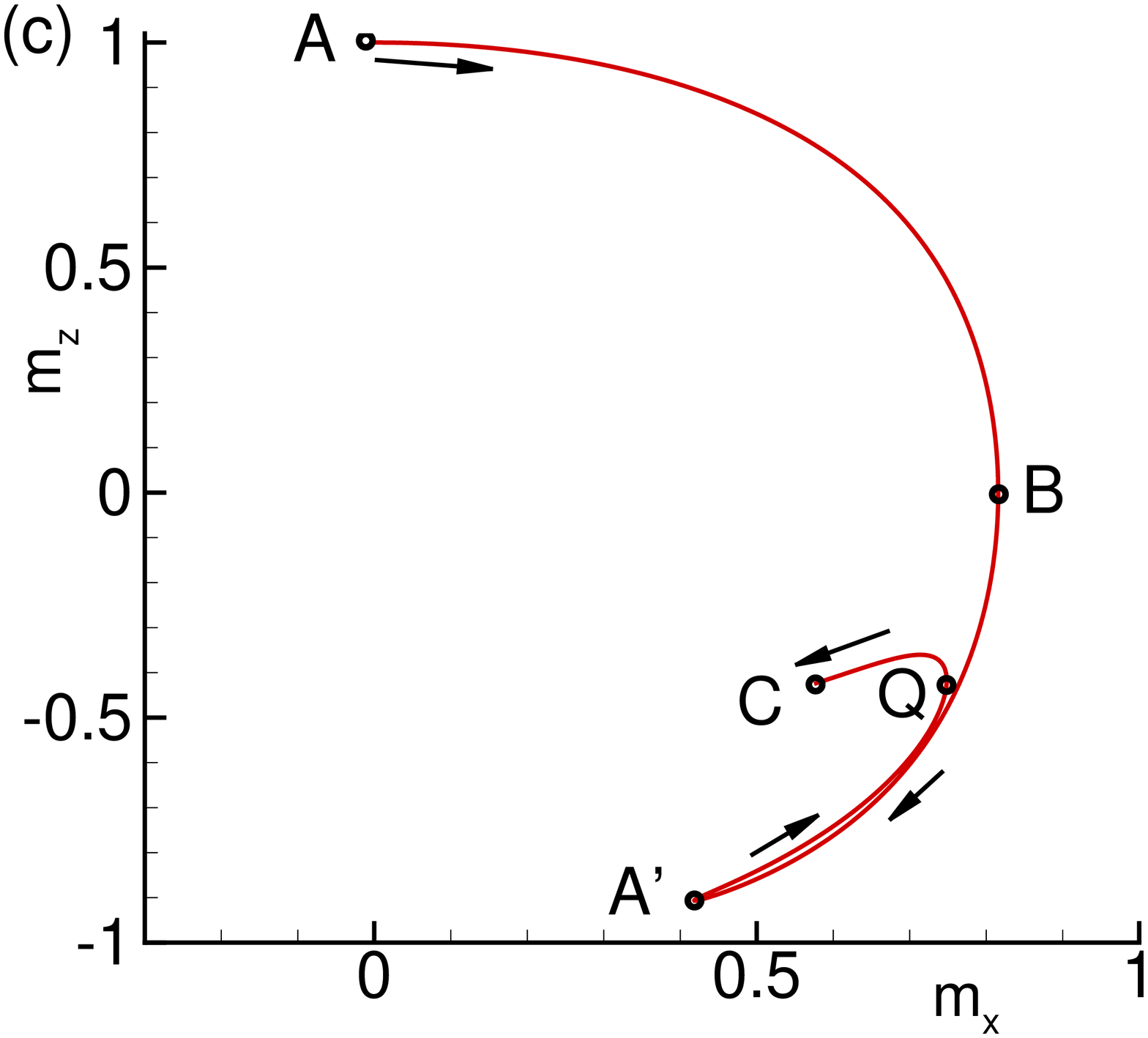} \includegraphics[height=35mm]{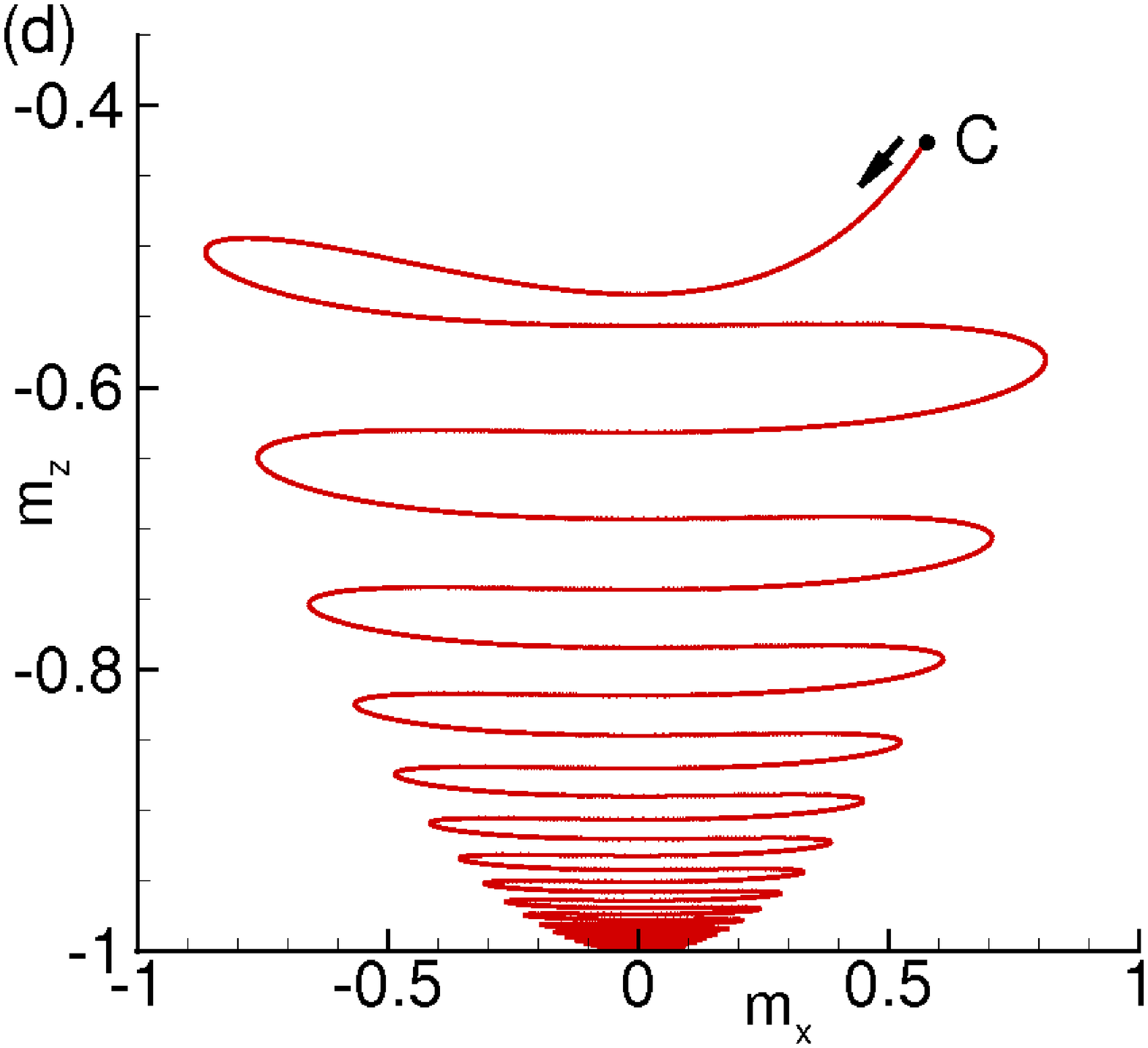}
 \caption{Trajectories of magnetization components in the planes $m_y-m_x$
 in the transition region:  (a) during electric pulse action (between points $A$ and $C$), (b) after switching the pulse off;  In (c) and (d)
 the same is shown for the $m_y-m_z$ plane. Parameters of the pulse
 and the JJ are the same as in Fig.\ref{2}(a) at $\alpha=0$. }
\label{3}
\end{figure}

The dynamics of the system in the form of magnetization trajectories in the planes $m_y-m_x$ and  $m_z-m_x$ during a transition time interval at the same parameters of the pulse and JJ at $\alpha=0$ is presented in Fig.\ref{3}.
We see that magnetic moment makes a spiral  rotation approaching the state with $m_z=-1$ after switching off the electric current pulse. The figures show clearly the specific features of the dynamics around points $B$, $A'$ and $Q$ and damped oscillations of the magnetization components (see Fig.\ref{3}(b) and Fig.\ref{3}(d)). The cusps at point B in Fig.\ref{3}(a) corresponds just to the change  from an increasing of absolute value of $m_x$ to its decreasing and, opposite, at point $A^{\prime}$ in Fig.\ref{3}(c). The behavior of magnetic system happens to be sensitive to the parameters of the electric current pulse and JJ. In the Supplement we show three additional protocols of the magnetization reversal by variation of $A_s$, $G$ and $r$.

\begin{figure}[h!]
 \centering
\includegraphics[height=50mm]{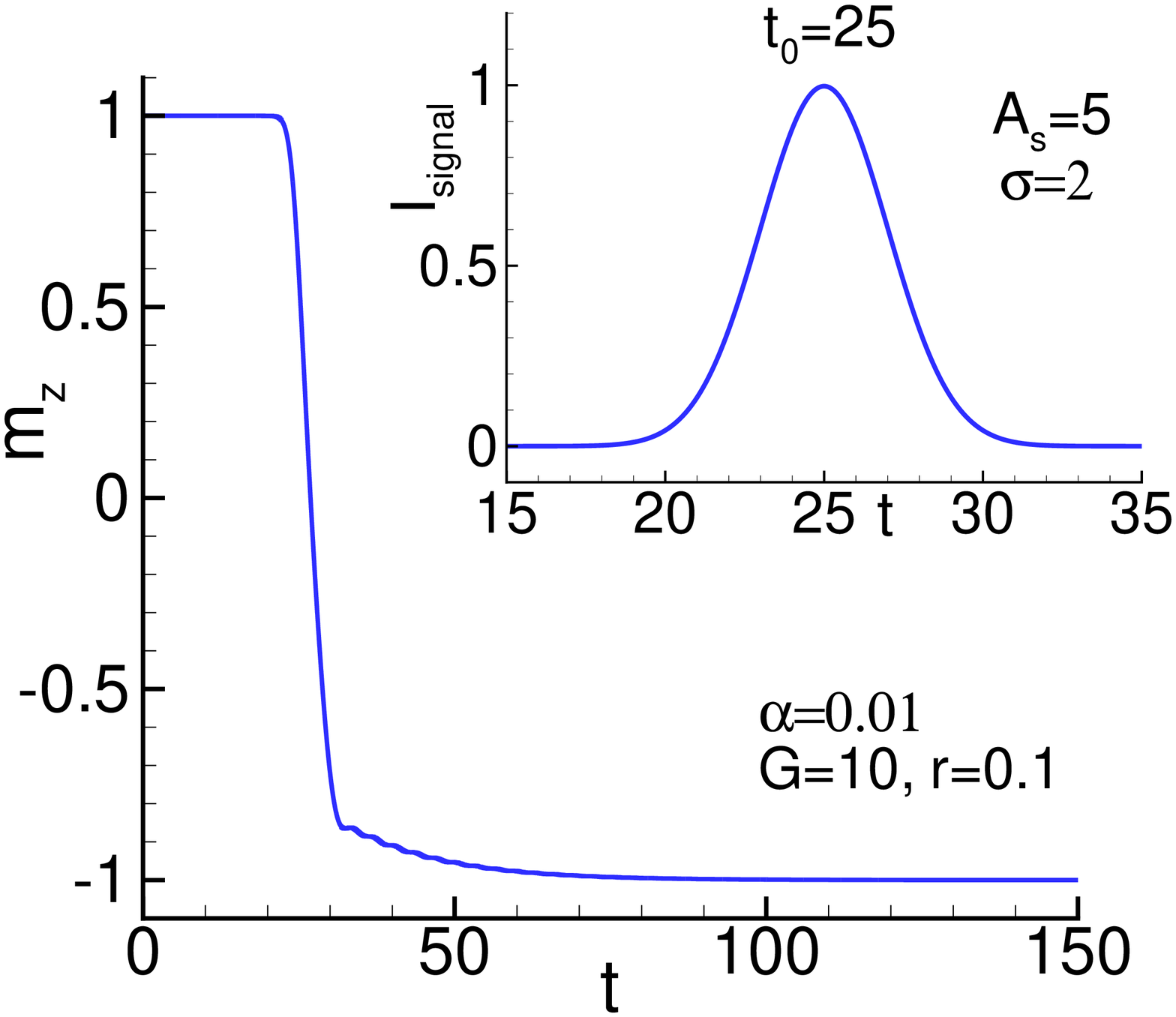}
\caption{Demonstration of transition dynamics of $m_z$ for a Gaussian electric current pulse (shown in the inset).}
\label{4}
\end{figure}

It is interesting to compare the effect of rectangular pulse with the Gaussian one of the form
\begin{equation}
\label{gauss1}
I_{pulse}=A_s\frac{1}{\sigma\sqrt{2\pi}}\exp\bigg(-\frac{(t-t_{0})^{2}}{2\sigma^{2}}\bigg).
\end{equation}
where $\sigma$ denotes the full width at half-maximum of the pulse
and $A$ is its maximum amplitude at $t=t_0$. In this case we also solve numerically the system of equations (\ref{syseq}) together with equation (\ref{current}) using (\ref{gauss1}).  An example of magnetic moment reversal in this case is presented in Figure \ref{4}, which shows the transition dynamics of $m_{z}$  for the parameters $r=0.1$, $G=10$, $A_s=5$, $\sigma=2$ at small dissipation $\alpha=0.01$.
We see that the magnetization reversal occurs more smoothly in compare with a rectangular case.

\begin{figure}[h!]
 \centering
 \includegraphics[height=35mm]{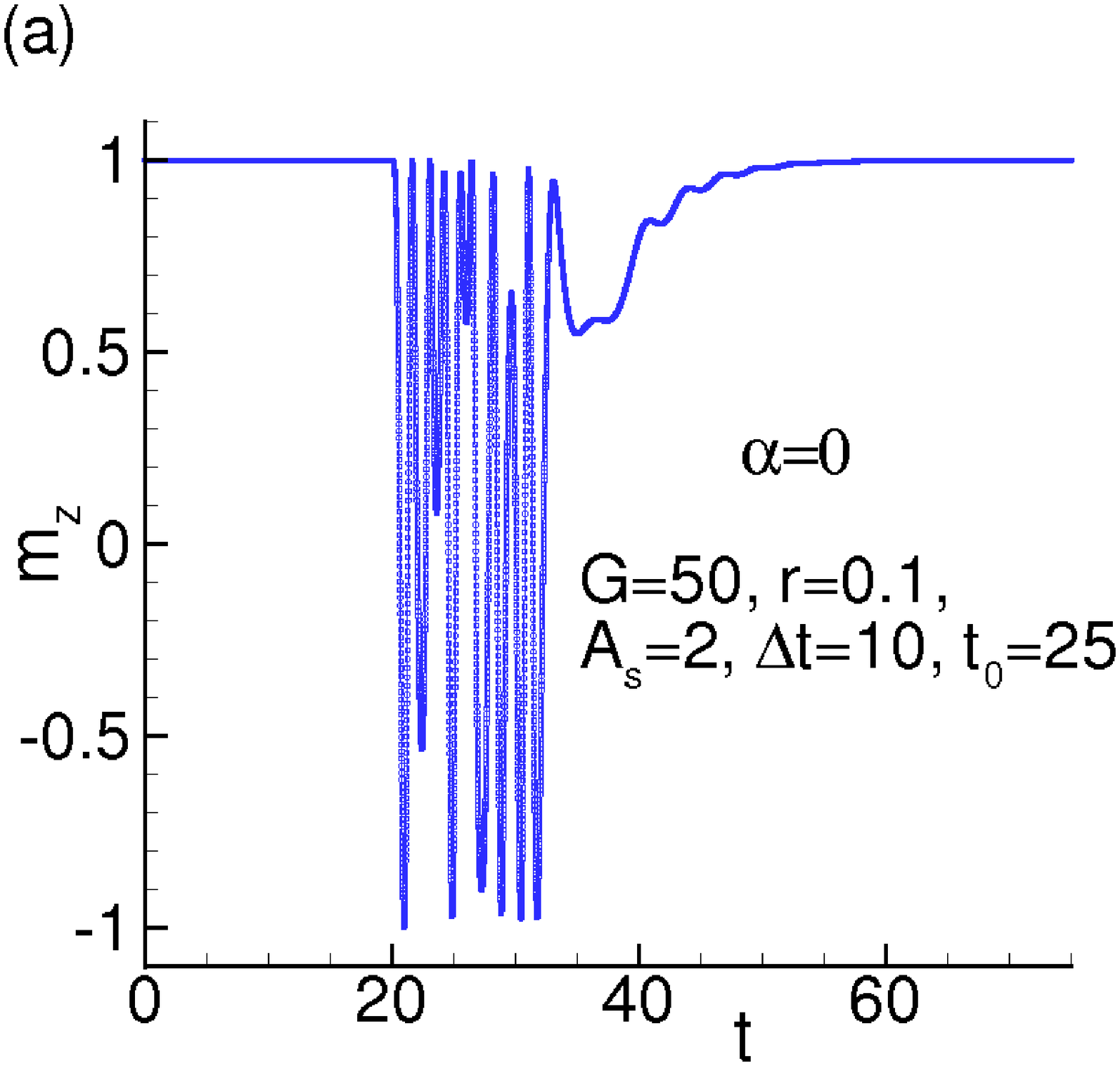}\includegraphics[height=35mm]{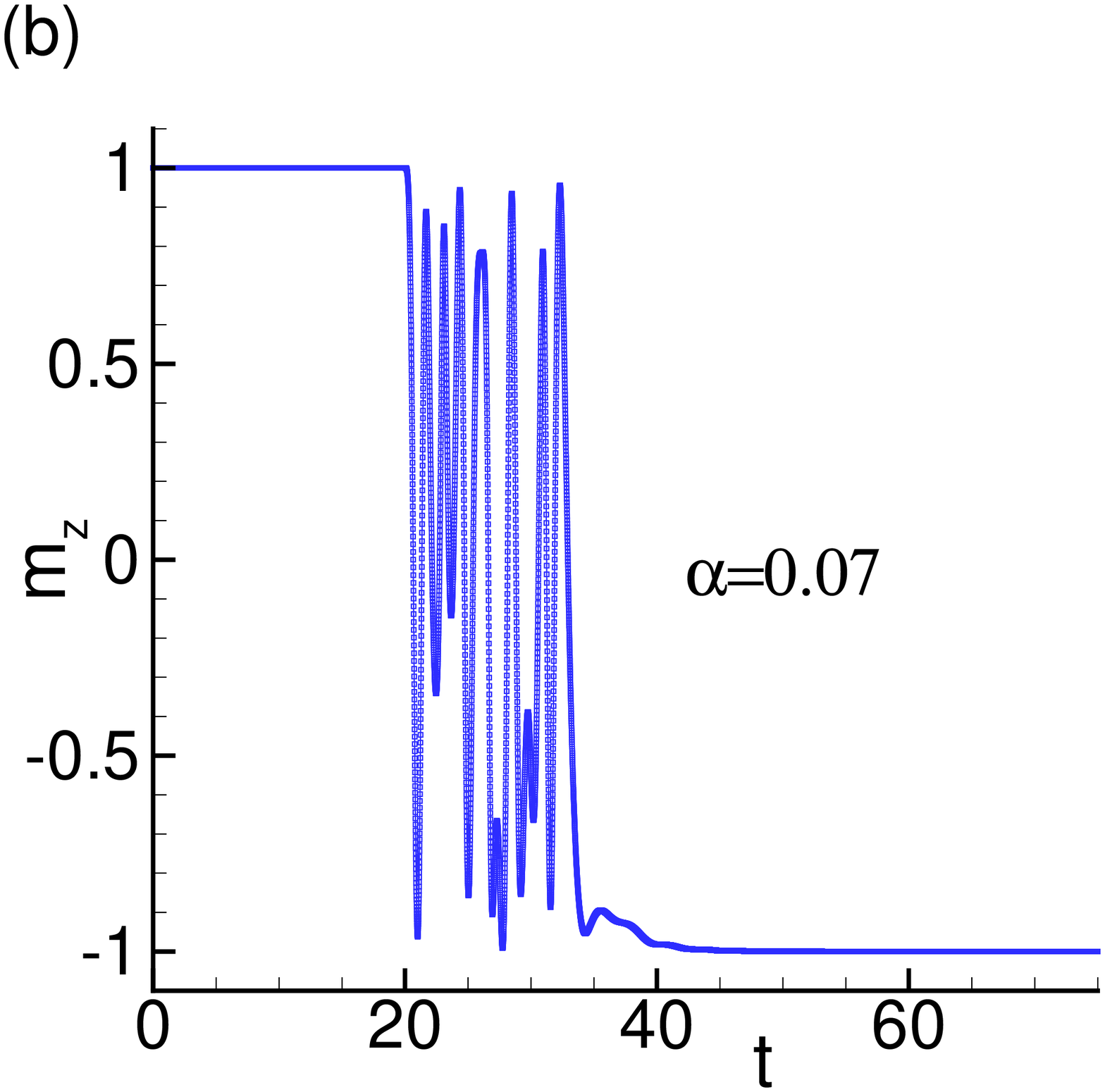}
 \caption{ Magnetization dynamics under rectangular pulse signal in the system at different values of the dissipation parameter $\alpha$.}
\label{5}
\end{figure}

\begin{figure}[h!]
 \centering
\includegraphics[height=50mm]{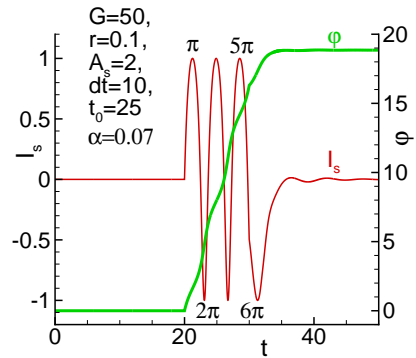}
\caption{Transition dynamics of the phase difference and the superconducting current for the case presented in Fig.\ref{5}(b).}
\label{6}
\end{figure}

We also note that very important role in the reversal phenomena belongs to the effect of damping. It's described by term with $\alpha$ in the system of equations (\ref{syseq}), where $\alpha$ is a damping parameter.  The examples of the magnetization reversal at $G=50$, $r=0.1$ and different values of $\alpha$ are presented in Fig.\ref{5}.
We see that dissipation can bring the magnetic system to full reversal, even if at $\alpha=0$ the system does not demonstrate reversal. Naturally, the magnetic moment, after reversal,  shows some residual oscillations as well. We stress that the full magnetization reversal is realized in some fixed intervals of dissipation parameter.
As expected, the variation of phase difference by $\pi$ reflects the maxima in the  time dependence of the superconducting current. Fig.\ref{6} demonstrates this fact.
The presented data shows that the total change of phase difference consists of $6\pi$, which corresponds to the six extrema in the dependence $I_s(t)$. After the full magnetization reversal is realized,  the phase difference shows the oscillations only.

One of the important aspect of the results that we obtain
here is the achievement of a relatively short switching time
interval for magnetization reversal. As we have seen in Figs.\
\ref{2}(a) and \ref{4}, the time taken for such reversal is
$\omega_F t \simeq 100$ which translate to $10^{-8}$ seconds for
typical $\omega_F \simeq 10 GHz$. We note that this amounts to a
switching time which is $1/20^{\rm th}$ of that obtained in Ref.\
\onlinecite{cai10}.

Experimental verification of our work would involve
measurement of $m_z(t)$ in a $\varphi_0$ junction subjected to a
current pulse. For appropriate pulse and junction parameters as
outlined in Figs.\ \ref{4} and \ref{5}, we predict
observation of reversal of $m_z$ at late times $\omega_F t \ge 50$.
Moreover, measurement of $m_y$ at times $t_{\rm max}$ and $t_{\rm
min}$ where $I_s$ reaches maximum and minimum values and the voltage
$V(t)$ across the junction between these times would allow for
experimental determination of $r$ via Eq.\ \ref{mageq1alt}.

As a ferromagnet we propose to use a very thin $F$ layer on dielectric substrate. Its presence produces the Rashba-type spin-orbit interaction and the strength of this interaction will be large in metal with large atomic number $Z$. The appropriate candidate is a permalloy doped with $Pt$.\cite{Hrabec} In $Pt$  the spin-orbit interaction play a very important role in electronic band formation and the parameter $\upsilon_{so}/\upsilon_{F}$, which characterizes the relative strength of the spin-orbit interaction is $\upsilon_{so}/\upsilon_{F} \sim 1$. On the other hand, the $Pt$ doping of permalloy up to 10 $\%$ did not influenced significantly its magnetic properties \cite{Hrabec}  and then we may expect to reach $\upsilon_{so}/\upsilon_{F}$ in this case 0.1 also. If the length of $F$ layer is of the order of the magnetic decaying length $\hbar \upsilon_{F}/h$, i.e. $l\sim 1$, we have $r\sim 0.1$.
Another suitable candidate may be a $Pt/Co$ bilayer, ferromagnet without inversion symmetry like MnSi or FeGe.
If the magnetic moment is oriented in plane of the F layer, than the spin-orbit interaction should generate a $\varphi_0$ Josephson junction\cite{buzdin08} with finite ground phase difference. The measurement of this phase difference (similar to the experiments in Ref.\onlinecite{Szombati}) may serve as an independent way for the parameter $r$ evaluation. The parameter $G$ has been evaluated in Ref.\onlinecite{konschelle09} for weak magnetic anisotropy of permalloy $K\sim 4\times 10^{-5}K\cdot {\AA}^{-3}$ (see Ref.\onlinecite{Rusanov}) and $S/F/S$ junction with $l\sim 1$ and $Tc\sim 10$K as $G\sim 100$. For stronger anisotropy we may expect $G\sim 1$.

In summary, we have studied the magnetization reversal in $\varphi_0$-junction with direct coupling between magnetic moment and Josephson current. By adding the electric current pulse, we have simulated the dynamics of magnetic moment components and demonstrate the full magnetization reversal at some parameters of the system and external signal. Particularly, time interval for magnetization reversal can be decreased by changing the amplitude of the signal and spin-orbit coupling. The observed features might find an application in different fields of superconducting spintronics. They can be considered as a fundamental basis for memory elements, also.

See supplementary material for demonstration of different protocols of the magnetization reversal by variation of Josephson junction and electric current pulse parameters.

{\it Acknowledgment}: The authors thank  I. Bobkova  and A.
Bobkov for helpful discussion. The reported study was funded by the
RFBR research project 16--52--45011$\_$India, 15--29--01217, the
the DST-RFBR grant INT/RUS/RFBR/P-249, and the French ANR projects
"SUPERTRONICS".

\newpage

\begin{widetext}
\begin{center}
    \LARGE{\textbf{Supplementary Material to ``Magnetization reversal by superconducting current in $\varphi_0$ Josephson junctions''}}
\end{center}
\end{widetext}

\section{ Geometry and equations}

\begin{figure}[h!]
 \centering
\includegraphics[height=60mm]{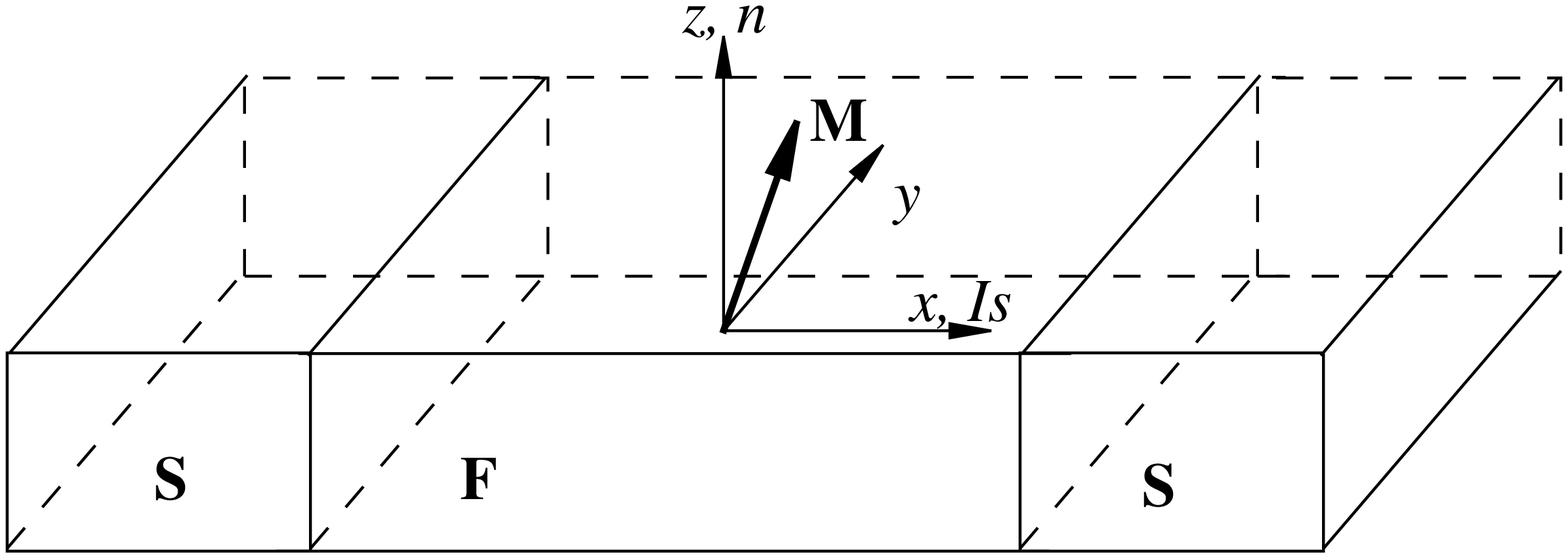}
\caption{ Geometry of the considered  $\varphi_0$-junction.}
\label{scheme}
\end{figure}

Geometry of the considered  $\varphi_0$-junction\cite{konschelle09} is presented in Fig.~\ref{scheme}. The ferromagnetic easy-axis is directed along the z-axis, which is also the direction n of the gradient of the spin-orbit potential. The magnetization component $m_y$ is coupled with Josephson current through the phase shift term $\varphi_0\sim (\overrightarrow{n}[\overrightarrow{m} \overrightarrow{\nabla} \Psi])$, where $\Psi$ is the superconducting order parameter ($\overrightarrow{\nabla} \Psi$ is along the x-axis in the system considered here).

In order to study the dynamics of the S/F/S system, we use the
method developed in Ref.\ \onlinecite{konschelle09}. We assume that
the gradient of the spin-orbit potential is along the easy axis of
magnetization taken to be along ${\hat z}$. The total energy of this
system is determined by the expression (S1) in the main text.

The magnetization dynamics is described by the
Landau-Lifshitz-Gilbert equation\cite{buzdin05}
\begin{equation}
\frac{d {\bf M}}{dt}=-\gamma {\bf M} \times {\bf
H_{eff}}+\frac{\alpha}{M_{0}}\bigg({\bf M}\times \frac{d {\bf
M}}{dt} \bigg) \label{momentum}
\end{equation}
where $\gamma$ is the gyromagnetic ratio, $\alpha$ is a
phenomenological Gilbert damping constant, and $M_{0}=\|{\bf M}\|$. The
effective field experienced by the magnetization ${\bf M}$ is
determined by $ \bf{H_{eff}}=-\frac{1}{\mathcal V}\frac{\partial E_{tot}}{\partial{\bf M}}$, so

\begin{eqnarray}
{\bf H_{eff}}&=
\frac{K}{M_{0}}\bigg[G r \sin\bigg(\varphi - r
\frac{M_{y}}{M_{0}} \bigg) {\bf\widehat{y}} +
\frac{M_{z}}{M_{0}}{\bf\widehat{z}}\bigg] \label{effective_field}
\end{eqnarray}

where $r=l\upsilon_{so}/\upsilon_{F}$, and $\displaystyle G=
E_{J}/(K \mathcal{V})$.

Using (\ref{momentum}) and (\ref{effective_field}), we obtain the system of equations (2) in main text, which describes the dynamics of the SFS structure.

\section{Magnetization reversal under electric current pulse}
Magnetic system is very sensitive to the parameters of the electric current pulse and Josephson junction. Here we show three additional protocols of the magnetization reversal by variation of $A_s$, $G$ and $r$.

\subsection{Effect of $A_s$-variation}
\begin{figure}[h!]
 \centering
\includegraphics[height=50mm]{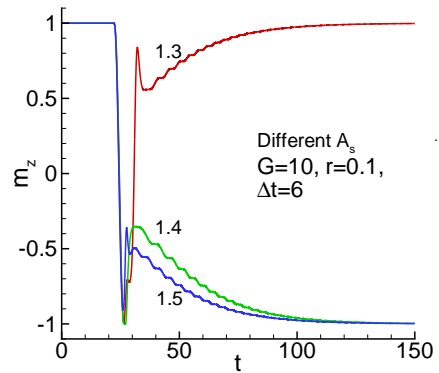}
\caption{Magnetization reversal by changing pulse parameter $A_s$. The number near curve shows value of $A_s$. }
\label{As}
\end{figure}

Figure ~\ref{As} demonstrates the magnetization reversal by changing pulse parameter $A_s$.
We see that change of pulse amplitude $A_s=1.3$ to $A_s=1.4$ reverses magnetic moment. At $A_s=1.5$ this feature is still conserved, but disappears at larger values.

\subsection{Effect of $G$-variation}
\begin{figure}[h!]
 \centering
\includegraphics[height=50mm]{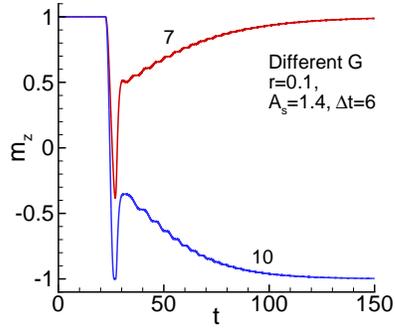}
\caption{Magnetization reversal by changing Josephson junction parameter $G$. The number near curve shows value of $G$.}
\label{G}
\end{figure}

Figure ~\ref{G} demonstrates the magnetization reversal by changing Josephson junction parameter $G$.

\subsection{Effect of $r$-variation}
\begin{figure}[h!]
 \centering
\includegraphics[height=50mm]{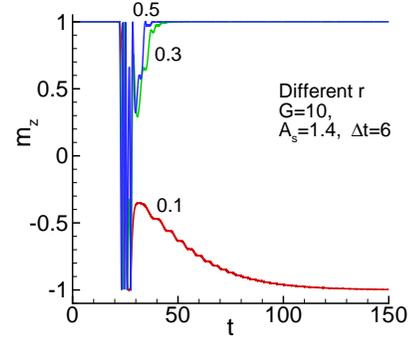}
\caption{Magnetization reversal by changing Josephson junction parameter of spin-orbital coupling $r$. The number near curve shows value of $G$.}
\label{r}
\end{figure}

Figure ~\ref{r} demonstrates the magnetization reversal by changing Josephson junction parameter of spin-orbital coupling $r$.
Figure ~\ref{r} demonstrates the magnetization reversal by changing Josephson junction parameter of spin-orbital coupling $r$.We see that there is a possibility of magnetization reversal around $G=10$. In this case a decrease of spin-orbit parameter may lead to the magnetization reversal also. The  magnetization reversal depends on the other parameters of the system  and, naturally, the minimal value of parameter  $r$ depends on their values. In particular case presented here it is around 0.05.

\end{document}